\documentclass[aps,prd,
twocolumn,amsmath,amssymb,amsfonts, long,preprintnumbers,superscriptaddress]{revtex4}
\usepackage{graphicx}

\renewcommand{\a}{a}
\def\g{g_{a\gamma}}
\newcommand{\Msun}{M_{\odot}}

\renewcommand{\gg}{g_{10}}

\newcommand{\m}{{m_\a}}

\newcommand{\E}{{\textbf{E}}}

\newcommand{\B}{{\textbf{B}}}

\newcommand{\GeV}{{\rm GeV}}

\newcommand{\mc}{\mathcal}

\begin{document}

\title{Probing the axion-photon coupling: phenomenological and experimental perspectives. 
A snowmass white paper}

\author{G. Carosi}
\email{carosi2@llnl.gov}
\affiliation{Lawrence Livermore National Laboratory, 7000 East Ave., Livermore, CA, 94550, USA}
\author{A. Friedland}
\email{friedland@lanl.gov}
\affiliation{Theoretical Division, T-2, MS B285, Los Alamos National Laboratory, Los Alamos, NM 87545, USA}
\author{M. Giannotti}
\email{mgiannotti@mail.barry.edu}
\affiliation{Physical Sciences, Barry University,
11300 NE 2nd Ave., Miami Shores, FL 33161, USA}
\author{M. J. Pivovaroff }
\email{pivovaroff1@llnl.gov}
\affiliation{Lawrence Livermore National Laboratory, 7000 East Ave., Livermore, CA, 94550, USA}
\author{J. Ruz}
\email{ruzarmendari1@llnl.gov}
\affiliation{Lawrence Livermore National Laboratory, 7000 East Ave., Livermore, CA, 94550, USA}
\author{J. K. Vogel}
\email{vogel9@llnl.gov}
\affiliation{Lawrence Livermore National Laboratory, 7000 East Ave., Livermore, CA, 94550, USA}

\date{September 25, 2013}

\preprint{LA-UR-13-27392}


\begin{abstract}

We present a brief overview of the ongoing searches for the axion particle via its coupling to photons. Both the classical QCD axions and more recently proposed Axion-Like-Particles are considered. Astrophysical bounds on the axion-photon coupling come from  considerations of stellar energy loss during Helium burning, in both low- and high-mass stars. Helioscopes look for 
back-conversion of solar axions into x-ray photons in strong laboratory magnetic fields. Finally, haloscopes aim to detect dark matter axions in our galactic halo. Both types of searches are expecting significant advances in the future, which will enable them to probe large, well-motivated parts of the parameter space below the stellar cooling bounds. 
\end{abstract}


\maketitle


\section{Introduction}

The experimental and theoretical investigation of the \textit{low energy frontier} of particle physics~\cite{Jaeckel:2010ni} has progressed considerably in the last years 
and new experimental proposals aim at probing large regions 
of the unexplored parameter space for WISPs (Weakly Interacting Slim Particles)
~\cite{Baker:2013zta}.

A particularly compelling case of a WISP is the axion~\cite{Weinberg:1977ma,Wilczek:1977pj}, 
a light pseudoscalar particle predicted by the most widely accepted solution of the strong CP problem~\cite{Peccei:1977hh,Peccei:1977ur} and the most prominent non-WIMP (Weakly Interacting Massive Particle) dark matter candidate~\cite{Abbott:1982af,Dine:1982ah,Preskill:1982cy}.

Axions are coupled to photons through the dimension 5 operator
\begin{equation}\label{Eq:agg}
	\mc L=-\frac{\g}{4}a F\tilde F=\g\,a \E\cdot\B
\end{equation}
and have mass
\begin{equation}
(m_{a}/1\mbox{ eV})= 0.5\, \xi \, \gg\,, 
\label{eq:massgA}
\end{equation}
where $ \gg=\g /(10^{-10}\GeV^{-1})$ and, in many motivated axion models, the dimensionless coefficient $\xi$ is of order 1. For example, for the well-known KSVZ \cite{Kim:1979if,Shifman:1979if} and DFSZ \cite{Dine:1981rt,Zhitnitsky:1980tq} scenarios we have $|\xi|\simeq 0.5$ and $1.4$ respectively.

Light pseudoscalar particles, weakly coupled to photons as in (\ref{Eq:agg}), emerge naturally in various extensions of the Standard Model (for recent reviews see \cite{Jaeckel:2010ni,Ringwald:2012hr,Baker:2013zta}). In recent years, a considerable attention was devoted to the so-called Axion-Like-Particles (ALPs), which couple to photons as in Eq. (\ref{Eq:agg}) but do not satisfy the mass-coupling relation (\ref{eq:massgA}). 

A series of unexplained astrophysical observations may provide additional motivation. 
Among these observations are the seeming transparency of the universe to very high energy gamma rays, the observed ratio of the blue to red Helium burning massive stars,
the larger than expected white dwarf cooling rates, and the quest for dark matter candidates.

The first of these problems concerns the propagation of high-energy gamma rays ($100~\GeV$ or more) in the galactic and extragalactic medium~\cite{Horns:2012fx,Albuquerque:2010rq}. 
In the standard model, these gamma rays should be damped by pair production on the background light, yet observations seem to indicate the that universe is more transparent to these photons than expected.
\emph{Photon-pseudoscalar} oscillations (in the extragalactic magnetic field) would provide an elegant solution to this transparency problem.
This mechanism requires a coupling in the range $ 0.1 \lesssim \gg \lesssim 1 $ and a mass well below the value predicted by Eq. (\ref{eq:massgA}), pointing, therefore, toward more general axion-like particles rather than the QCD axions \cite{DeAngelis:2011id,Simet:2007sa}.

Another possible astrophysical hint comes from the observations of the numbers of blue versus red supergiants. 
It has been reported that there are fewer blue stars than predicted by stellar evolution models (see~\cite{McQuinn:2011bb} and references therein). 
These blue stars appear also to be less blue than expected.
To properly gauge the significance of this effect, a more complete analysis of stellar models is required, but it is intriguing that an axion (or ALP) with a coupling to photons in the same range as required by the transparency problem would alleviate both issues. It should also be mentioned that this parameter range can be explored by the next generation axion helioscopes (see sec.~\ref{sec:IAXO}).

The existence of WISPs is also hinted at by the analysis of the white dwarf luminosity function and the recently measured decrease of the pulsation period of ZZ-Ceti G117-B15A~\cite{Isern:2008fs,Ringwald:2012hr}. 
Observations seem to indicate an anomalously large cooling rate, which can be explained by the emission of axions or ALPs produced by electron bremsstrahlung in the stellar interior~\cite{Isern:2012ef,Isern:2013lwa}.
For a standard axion, the required axion-electron interaction strength would also imply an axion-photon coupling, however in a range about an order of magnitude lower than what indicated by the previous problems.

Last, but certainly not least, the axion provides an excellent dark matter candidate. This possibility has been actively studied for three decades.
In fact, the recent analysis of the galactic angular momentum distribution seems to indicate a preference of axions over WIMPs as cold dark matter (CDM) candidates and that axions are likely to be at least a component of the total dark matter in the galaxy~\cite{Banik:2013rxa}.
The searches for axions in our galactic halo are summarized in sec.~\ref{sec:Haloscopes} below.

Apart from hints, astrophysical considerations have provided strong limits on the allowed range of the axion-photon coupling. It is well-known~\cite{PDG2010} that accurate counts of the Horizontal Branch (HB) stars provide a powerful constraint, $ \gg \lesssim 1 $~\cite{Raffelt:1987yu}.
Moreover, as has been recently noticed, the analysis of the evolution of massive stars has led to the bound $ \gg \leq 0.8 $~\cite{Friedland:2012hj}. The bound is based on the fact that Cepheid variable stars exist, as described below in sec.~\ref{sec:blueloop}.

Several experiments are currently involved in the axion/ALP search.
Here, we focus on three experiments in particular which are probing  regions of the axion-photon coupling in the mass range $ \m<$ a few eV: the CERN Axion Solar Telescope (CAST), based at CERN (see sec.~\ref{sec:CAST}), the Axion Dark Matter eXperiment (ADMX), based at the U. of Washington (see sec.~\ref{sec:ADMX}), and the newly funded Axion Dark Matter eXperiment High-Frequency (ADMX-HF) (see sec.~\ref{sec:ADMX-HF}), based at Yale U.
The last two experiments are searching for dark matter axions in the galactic halo while the first is an example of axion helioscope and it is looking for axions from the Sun.

Recently there has been a proposal for a \textit{next generation axion helioscope}, the International Axion Observatory (IAXO) (see sec.~\ref{sec:IAXO}), with the potential of exploring a large range of the axion-photon coupling, more than an order of magnitude below the current CAST bound in $ \g $. 
In addition, IAXO will partly probe axion and ALP CDM parameter region.

All these experiments involve the conversion of axions into photons in an external magnetic field \cite{Sikivie:1983ip}. 
A notorious difficulty of this technique is the fact that the conversion probability of axions into photons in an external magnetic field is strongly dependent on the axion mass and this makes it harder to probe large regions of the coupling-mass parameter space. 
This was a major motivation for CAST's $^{4}$He/$^{3}$He phases as well as for ADMX-HF to push on the mass in addition to the coupling front.

The effort of studying the axion-photon coupling is supported in the United States by grants from the National Science Foundation (NSF) and the Department of Energy's Office of Science (DOE/SC) that support ADMX (DOE/SC) and ADMX-HF.   Laboratory Directed Research and Development (LDRD) programs at Lawrence Livermore National Laboratory (LLNL) and Los Alamos National Laboratory (LANL) have supported work in CAST (LLNL) and in phenomenological analyses (LANL). 

In this report, as a contribution to the Snowmass Cosmic Frontier planning process, we present a summary and perspective on the experimental and astrophysical probing of the axion-photon couping.
In particular, in section \ref{sec:astro} we will discuss the new phenomenological bound from the analysis of the evolution of massive stars; in section~\ref{sec:helioscopes} the 
status of the axion helioscope experiment CAST and the perspective on the new IAXO proposal; and in section~\ref{sec:Haloscopes} the axion dark matter experiments ADMX and ADMX-HF. 
In the conclusion, we will summarize the possible perspectives in this field and discuss possible motivations for this effort. 

\section{Bounds from stellar evolution}
\label{sec:astro}

Stars are efficient laboratories to study the properties of \emph{light, weakly interacting} particles, more specifically, particles that are light enough to be easily thermally produced in the stellar core and have a sufficiently large mean free path in the stellar matter to 
easily escape. 
 
For stars in the early evolutionary stages, where observations are abundant, 
a particle is \emph{light} if its mass does not exceed a few keV (for main sequence stars) to tens of keV (for red giants and He-burning stars). 
These values of the mass are actually fairly high by the standards of terrestrial experiments. In certain regions of the parameter space, stellar constraints dominate over the terrestrial bounds by several orders of magnitude (see the following sections).

Stars in later evolutionary stages (C-burning and beyond) have higher core temperatures but the observations in these cases are less accurate and effective bounds are rarer (see, however, \cite{Heger:2008er}).

Observations require that the energy drain be not too different from the Standard Model (SM) predictions~\cite{Raffeltbook}. 
Therefore, a novel process has a chance to be constrained by stellar evolutionary arguments if it contributes to the energy loss rate in an amount similar to the predictions of the standard model. 

Light axions or ALPs coupled to photons as in (\ref{Eq:agg}) would be produced in the stellar core through the Primakoff process~\cite{Primakoff51}, that is by converting thermal photons into axions in the plasma electromagnetic field~\cite{Dicus:1978fp,Fukugita:1982gn,Raffelt:1985nk}.
The production rate in a nondegenerate plasma is a well established result~\cite{Raffelt:1985nk,Raffeltbook}
\begin{equation}
\epsilon_{A}= Z(\xi^{2})\frac{g_{a\gamma}^{2}}{4\pi^{2}} \frac{T^{7}}{ \rho} 
\simeq 28 \frac{\mbox{erg}}{\mbox{g} \cdot \mbox{s} } Z(\xi^{2}) g_{10}^{2} T_{8}^{7} \rho_{4}^{-1}, 
\label{eq:loss}
\end{equation}
where $\rho_4\equiv\rho/(10^4$~g/cm$^{3})$, $T_8 \equiv T/10^8 \rm K$ and $Z(\xi^{2})$ is a function of $\xi^{2}\equiv(\kappa_{S}/2T)^{2}$, with $\kappa_{S}$ being the Debye-Huckel screening wavenumber.
$Z(\xi^{2})$ is given as an integral over the photon distribution (see Eq.~(4.79) in \cite{Raffelt:1990yz}) and is generally ${\cal O}(1)$ for relevant stellar conditions.
An accurate analytical approximation of $Z(\xi^{2})$ can be found in~\cite{Friedland:2012hj}.

A comparison of the emission rate with the standard energy loss allows to define the density and temperature conditions in which axion cooling would be most relevant. 
This region corresponds to 
moderately high temperatures ($ T\sim 1-2\times 10^8 $ K) and relatively low densities 
($ \rho =10^3-10^4 \mbox{ g} \cdot \mbox{cm}^{-3}$), 
typical of the core of He-burning stars.

It is, in fact, from the phenomenology of stars in the He-burning phase that the best constraints on $  \g  $ are derived, as we discuss below.

\subsection{Low mass stars}

A restrictive constraint on $ \gg $ arises from low-mass, $ M\sim 1 \Msun $, Horizontal Branch (HB) stars in globular clusters~\cite{Raffelt:1987yu}.
A recent review of this bound can be found in section 3.4.1 of ~\cite{Kuster:2008zz}.

These stars have He-burning core of density $ \rho \simeq 10^4 \mbox{ g} \cdot \mbox{cm}^{-3} $ and temperature $ T\simeq 10^8 $ K.
The main effect of the axion production in such stars would be the reduction of the helium burning lifetime.
This would have the observable consequence of reducing the number of HB versus red giant (the stage that precedes the He burning phase) stars in globular clusters. 

A numerical simulation of the evolution of a 1.3 $ \Msun $ star of solar metallicity~\cite{Raffelt:1987yu} showed that a value of $ \gg=1 $ causes a reduction of the He-burning time by about 30\%, a value considered observationally forbidden ({\it cf.}~\cite{Kuster:2008zz}, section 3.4.1).

The bound $ \gg<1 $, shown in Figure~\ref{Fig:overview} below, extends up to 30 keV or so in the ALP mass~\cite{Raffeltbook} and is strongly suppressed at higher masses (well beyond the region shown in the figure).
 

\subsection{Massive stars}
\label{sec:blueloop}
\begin{figure}[bt]
  \includegraphics[angle=0,width=0.88\columnwidth]{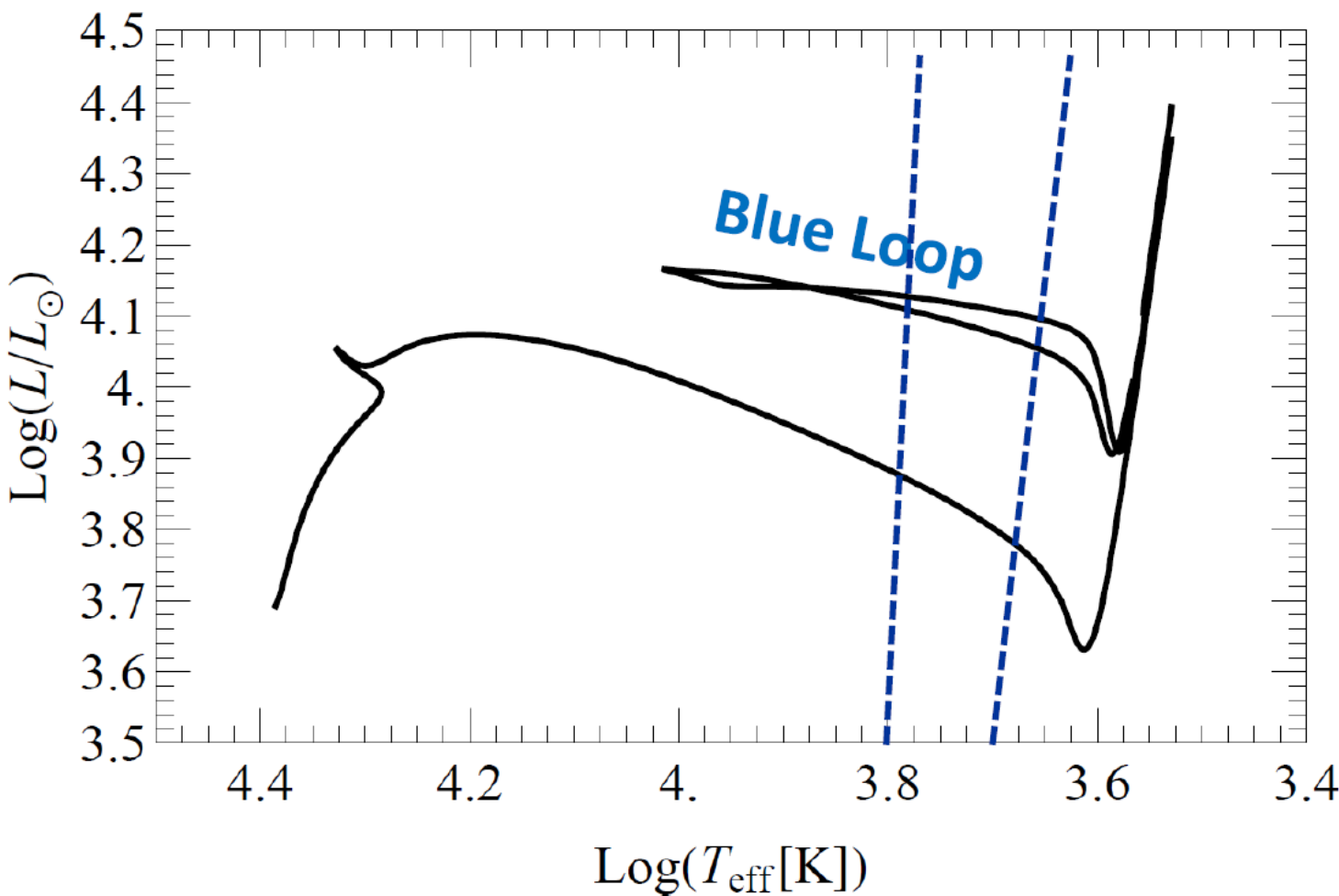}
  \caption{Evolution in the HR diagram of a $ 9.5~M_{\odot} $ star. The region between the two dashed lines is the instability strip, where the Cepheid variables are found.}\label{Fig:BlueLoop}
\end{figure}
A somewhat stronger bound can be derived by the analysis of the evolution of massive stars~\cite{Friedland:2012hj}.
These stars, while burning helium in their core, evolve into a stage called the \emph{blue loop} during which they contract and expand again.
In the Hertzsprung-Russell (HR) diagram, this appears as a loop which extends toward the left, 
(see Figure \ref{Fig:BlueLoop}), 
in the region of higher surface temperature (\emph{blue}) before turning back in the colder (\emph{red}) region on the right side of the diagram.

Recent stellar modeling efforts 
showed that the additional cooling  induced by axions coupled to photons would eliminate this evolutionary stage for stars of mass between 8 and 12 $ M_{\odot} $, unless $  g_{10} < 0.8  $~\cite{Friedland:2012hj}. 
This is the strongest bound to date on the ALP-photon cooling for masses below a few 10 keV~\cite{Raffelt:2013mea}. 

Interestingly, in this case the main observational effect of the nonstandard cooling is not the reduction of the He-burning lifetime but the disappearance of an evolutionary stage. 
This provides, therefore, a qualitative change in the stellar evolution.
The existence of the loop is corroborated by many astronomical observations. 
In particular, this stage is essential to account for the observed Cepheid stars
(see, e.g., \cite{kippenhahn}).
The complete disappearance of the  blue loop stage, even for a small range of luminosities (stellar masses), is observationally forbidden.

\begin{figure}[bt]
  \includegraphics[angle=0,width=0.88\columnwidth]{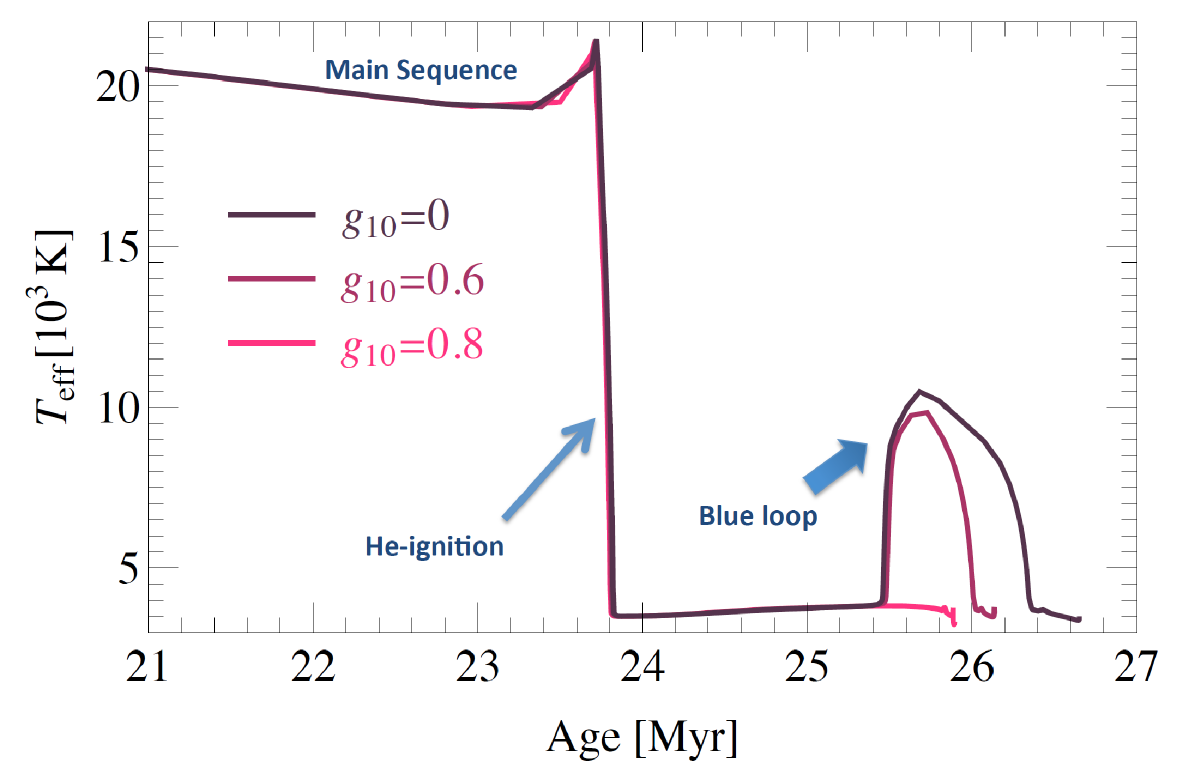}
  \caption{Impact of $g_{a\gamma}$ on the evolution of a 9.5 $\Msun$ star through the $He$ burning stage. Shown is the evolution of the surface temperature, for the cases $g_{10}=0$, $g_{10}=0.6$, and $g_{10}=0.8$. Axion losses speed up $He$ burning, resulting in the shortening and eventual elimination of the blue loop stage.}
\label{Fig:Teff}
\end{figure}

Lower values of $ \g $ would reduce the time a star spends in the blue loop stage, as shown in Fig.~\ref{Fig:Teff} and, consequently, the number of \emph{blue} versus \emph{red} stars of a given luminosity. 
An accurate observational analysis (see, e.g., \cite{McQuinn:2011bb} for a recent discussion on the status of the observations) could, therefore, potentially provide a more restrictive bound. Conversely, a persisting disagreement between the observations of the blue helium burning stars and stellar models, as mentioned in the introduction, could indicate the presence of the axion.  


%
\section{Helioscopes}
\label{sec:helioscopes}
%
%
As discussed above, thermal production of axions is expected to take place in the interiors of stars. 
Via the Primakoff effect, blackbody photons in the solar core can be converted into axions due to the presence of the strong electromagnetic fields in the plasma. These extraordinarily weakly coupled particles then escape the Sun and stream largely undisturbed to Earth, where experiments can make use of the inverse Primakoff effect in order to backconvert them into x-ray photons inside a powerful transverse magnetic field~\cite{Sikivie:1983ip,Sikivie:1985yu,vanBibber:1988ge}. 

The axion-photon conversion probability in a magnetic field filled with a buffer gas is given as
\begin{equation}
P_{a\rightarrow\gamma}= 
\left(\frac{Bg_{a\gamma}}{2}\right)^{2} \frac{1+e^{-\Gamma L}-2e^{-\Gamma L/2}\cos(qL)}{q^{2}+\Gamma^{2}/4}
\end{equation}
 where where $B$ and $L$ are the magnetic field strength and length, respectively, $\Gamma$ is the inverse photon absorption length in the
buffer gas, and the axion-photon momentum transfer is $q=\left| m_{a}^{2}-m_{\gamma}^{2}\right| /{2E}$.

In order to search for solar axions with high sensitivity such an experiment, dubbed an axion helioscope, needs to be composed of a powerful magnet of large volume along with low-background, high-efficiency x-ray detectors coupled to the end of the magnetic field region. X-ray optics can be used to enhance the discovery potential of a helioscope by focusing the potential signal from axion conversion and hence reducing detector size and ultimately background levels. Ideally, the magnet is equipped with a precise azimuth and horizontal drive to maximize the exposure time of the experiment by keeping it aligned with the solar core for the longest time possible.

\subsection{The CERN Axion Solar Telescope: CAST }
\label{sec:CAST}
The first axion helioscope operated at Brookhaven National Laboratory in 1992 using a static dipole magnet \cite{Lazarus:1992ry}, while a second-generation experiment, the Tokyo Axion Helioscope, carried on with the quest employing a more powerful magnet ($4$~T) and dynamic tracking of the Sun \cite{Moriyama:1998kd,Inoue:2002qy,Inoue:2008zp}. The CERN Axion Solar Telescope (CAST), a helioscope of the third generation and the most sensitive solar axion search to date, began data collection in 2003~\cite{Zioutas:1998cc}. It employs an LHC dipole test magnet of 10~m length and 10~T field strength with an elaborate elevation and azimuth drive to track the Sun. CAST is the first solar axion search exploiting x-ray optics to improve the signal to background ratio (a factor of 150 in the case of CAST~\cite{Kuster:2007ue}). No signal above background was observed, but CAST set an upper limit of $g_{a \gamma} < 8.8\times 10^{-11}$~GeV$^{-1}$ for $m_{a} <  0.02$~eV (vacuum phase,\cite{Zioutas:2004hi,Andriamonje:2007ew}), and a slightly larger value of $g_{a \gamma}$ for higher axion masses ($^{4}$He/$^{3}$He phase, \cite{Arik:2008mq,Arik:2011rx,Arik:2013nya}), exploring previously untested axion model space and probing the KSVZ model for the first time at the upper end of the mass range. Recently, CAST has also established the first helioscope limits for non-hadronic axion models~\cite{Barth:2013sma}. Currently, the experiment is refining its search by revisiting its $^{4}$He and vacuum phase with enhanced detectors and improved sensitivity to extend testing of the KSVZ model as well as ALP motivated parameter regions. 

\subsection{Future Prospects of Solar Axion Searches: IAXO }
\label{sec:IAXO}
To date, all axion helioscopes have used ``recycled'' magnets built for other purposes and each subsequent generation of axion helioscopes has resulted in an improvement in sensitivity to the axion-photon coupling constant $g_{a \gamma}$ of about a factor 6 over its predecessors. A larger and more sensitive, $4^{\rm th}$ generation helioscope, the International Axion Observatory (IAXO,~\cite{Irastorza:2011gs}), has recently been proposed. Significant conceptual design work has been completed and a Letter of Intent solicited by CERN was submitted~\cite{Irastorza13}. IAXO's improved sensitivity relies on a purpose-built large-scale magnet, efficient x-ray focusing optics and low-background detectors and aims to push the existing CAST sensitivity towards lower couplings over a large range of axion masses. 
 
The primary physics goal for IAXO is to search for axions or ALPs produced in the Sun via Primakoff conversion of solar plasma photons and to substantially improve on all previous solar searches. IAXO will have about $4-5$ orders of magnitude better sensitivity in terms of signal-to-background ratio than CAST translating into a factor of $\approx 20$ in terms of the axion-to-photon coupling constant $g_{a\gamma}$. This will allow the instrument to reach the $g_{a\gamma}$-regime of a few $\times 10^{-12}$~GeV$^{-1}$ for a wide range of axion masses up to about $0.25$ eV. IAXO will therefore enter into completely unexplored axion and ALP parameter space. Needless to say that the discovery of a new pseudoscalar would be a groundbreaking result for particle physics, but even in case of non-detection, IAXO will be able to exclude a large region of the unexplored QCD axion parameter space.

More specifically, the experiment would be able to explore a broad range of realistic QCD axion models at high mass end of the parameter region. Its sensitivity would cover theoretically favored model space with masses down to a few meV, superseding the SN 1987A energy loss limits ($m_{a} < 16$~meV for hadronic axions) and probing the axion hint from anomalous white dwarf cooling. Axion models in this region and below are of high cosmological interest as favored dark matter candidates and they could compose all or part of the cold dark matter of the Universe. In non-standard cosmological scenarios, more generic ALP frameworks~\cite{Arias:2012az} or mixed axion-WIMP DM scenarios~\cite{Baer:2011hx}, the range of axion and ALP parameters of interest as CDM is enlarged and most of the region at reach for IAXO contains possible dark matter candidates. At much lower masses, below $\approx 10^{-7}$~eV, the region explorable by IAXO includes ALP parameters invoked repeatedly to explain anomalies in light propagation over astronomical distances~\cite{DeAngelis:2011id,Horns:2012fx,Simet:2007sa,Albuquerque:2010rq}. With IAXO a definitive test of this hypothesis would be provided.

Additional physics cases for IAXO include the possibility of detecting more specific models of axions or ALPs from the Sun. Most noteworthy is the possibility to detect the flux of solar axions produced by axion-electron coupling $g_{ae}$ induced phenomena, including bremsstrahlung, Compton scattering, and axio-recombination. Although the existence of these production channels for standard axions is model-dependent, axions with a $g_{ae}$ of few $\approx 10^{-13}$ have been invoked to solve the anomalous cooling observed in white dwarfs. Similarly IAXO will be sensitive to models of other proposed particles like hidden photons or chameleons~\cite{Brax:2010xq}, scalars with an environment-dependent mass proposed in the context of dark energy models. The possibility of directly testing the particle physics nature of dark energy is an exciting prospect, even though this is still at an early stage of theoretical development.

Further potential experimental programs for IAXO may include a search for axionic dark radiation~\cite{Conlon:2013isa}, the realization of microwave LSW experiments among different bores of the IAXO magnet~\cite{Betz:2012tp}, and the direct detection of relic CDM axions~\cite{Hiramatsu:2012gg} or ALPs using microwave cavities or antennas in different configurations within the IAXO magnet. While the physics potential of all these options is currently under study, they certainly offer IAXO the opportunity to grow into a versatile, first-class multi-purpose facility for axion research.

\section{Dark Matter Haloscopes}
\label{sec:Haloscopes}

The first tractable experiment to find light dark matter axions in the $\mu$eV mass range was suggested by Sikivie in 1983 \cite{Sikivie:1983ip}. It relies on the Primakoff effect where the axion converts it's rest mass into a detectable photon in the presence of a static magnetic field. By placing a resonant microwave cavity in the bore of a strong solenoid magnet the axion production rate can be greatly enhanced when the resonant frequency of the cavity matches the mass of the axion. These converted photons can then be distinguished  from background by using ultrasensitive cryogenic microwave detectors. The power generated in the haloscope when the axion mass is on resonance goes as:
\begin{equation}
\label{equ:power}
P_a = g^2_{a\gamma} V B^2_0 C_{mnl} \left(\frac{\rho_a}{m_a}\right) {\rm min}(Q_L,Q_a).
\end{equation}
Here $V$ is the volume of the cavity, $B_0$ is the magnetic field, $\rho_a$ is the local dark matter axion mass density, $m_a$ is the axion mass, $Q_L$ is the loaded quality factor of the cavity, $Q_a \approx 10^6$ is the axion quality factor (ratio of axion energy over the energy spread), and $C_{mnl}$ is a mode dependent form factor given by
\begin{equation}
\label{equ:form_factor}
C_{mnl} = \frac{|\int_V d^3x \vec{E}_{\omega}\cdot\vec{B}_0|^2}{B^2_0 V \int_V d^3x \epsilon |\vec{E}_{\omega}|^2}
\end{equation}
where $\vec{B}_0$ is the static magnetic field, $\vec{E}_{\omega}$ is the oscillating electric field of the cavity mode, and $\epsilon$ is the dielectric constant \cite{Sikivie:1985yu}. For an empty cylindrical cavity in a homogenous magnetic field the largest form factor applies to the $TM_{010}$ mode with $C_{010} \sim 0.69$. Higher $TM_{0n0}$ modes have much smaller form factors and the remaining TM modes (along with the TE and TEM modes) integrate to form factors of 0.
Since the axion mass is unknown the cavity is tuned over a large range of frequencies during a search. This is accomplished by moving dielectric or metallic tuning rods inside the cavity. The speed in which a mass range can be scanned for axions at a given signal-to-noise ratio (SNR), or to exclude them at a particular confidence level, is guided by the Dicke radiometer equation:
\begin{equation}
\label{equ:dicke}
SNR = \frac{P_a}{P_N}\sqrt{B t} = \frac{P_a}{k_B T_S}\sqrt{\frac{t}{B}}
\end{equation}  
where $P_N = k_B B T_S$ is the thermal noise power, $B$ is the bandwidth and $T_S$ is the system noise temperature (physical + receiver noise) \cite{Dicke:1946}. 

First generation prototype experiments were carried out at Brookhaven National Laboratory \cite{DePanfilis:1987} and the U. of Florida \cite{Hagmann:1990} and were able to get within two orders of magnitude of the KSVZ limit~\cite{Kim:1979if,Shifman:1979if}.  Both of these experiments, along with the ones listed in more detail below, used linear amplifiers for signal power detection. An alternative method of detecting microwave photon employing rydberg atoms as single photon counters was tested by the CARRACK experiment at Kyoto U. of Japan \cite{Tada:2006}. 

\subsection{The Axion Dark Matter Experiment (ADMX)}
\label{sec:ADMX}

The Axion Dark Matter Experiment (ADMX) is a large-scale haloscope that was constructed in the early 1990s at LLNL. It is comprised of a collaboration of LLNL, U. of Washington, U. of Florida, U. of California at Berkeley, the National Radio Astronomy Observatory (NRAO), and the U. of Sheffield in the U.K. It consists of an 8 Tesla superconducting solenoid magnet with a 0.5 meter diameter, 1 meter long bore. The microwave cavity consist of a stainless steel right-circular cylinder coated in OFHC copper and annealed which, at cryogenic temperatures, have loaded quality factors around $Q_L \sim 10^5$. Two metallic (or dielectric) tuning rods are used to raise (or lower) the resonant frequency of the cavity as it searches for the axion. 

Power generated in the microwave cavity is coupled to a cryogenic receiver chain by a coaxial antenna. A second weakly coupled antenna is used to inject swept signals to determine the resonant frequency of the cavity at any particular moment. The power generated in the cavity is then amplified by several stages of cryogenic  and room-temperature amplifiers before the signal is mixed down to audio frequencies. Here the data is passed to a FFT spectrum analyzer which records the power-spectrum. Axion signals would show up as excess power above the noise floor. The power generated by axions should scale as $B^2$ giving a relatively simple discrimination to other potential external sources. 

ADMX has gone through three major phases. The first experimental runs used pumped liquid helium to cool the system to $T_{phys} \approx 2 K$ and used balanced GaAs HFET amplifiers supplied by NRAO with noise temperatures of $T_N \approx 2 K$ leading to a system noise temperature of $T_{sys} = T_{phys} + T_N \approx 4 K$ \cite{Bradley:2003,Asztalos:2004}. The next phase saw the introduction of SQUID amplifiers, which can scale to much lower noise temperatures than the HFETs, and can get close to the quantum limit (defined as $T_Q \sim h\nu/k$). An axion search was performed with SQUID amplifiers from 2008-2010 at 2 K \cite{Asztalos:2010,Asztalos:2011}. Subsequently the experimental apparatus was moved to the U. of Washington where it is currently being upgraded to operate at much lower physical temperatures (initially for 9-12 months with a $^3$He pot at $T_{phys} \approx 400\;mK$ and then with a dilution refrigerator at $T_{phys} \approx 100\;mK$). This will greatly increase the scan rate of and allow for searches for axions at even pessimistic DFSZ coupling sensitivity. The initial data runs will search frequency ranges from 0.5 - 2 GHz (2-8 $\mu$eV) and research is currently ongoing to increase that reach to higher (and lower) frequencies with the development of new cavity structures and amplifiers. 

\subsection{The Axion Dark Matter Experiment High Frequency (ADMX-HF)}
\label{sec:ADMX-HF}

The Axion Dark Matter Experiment High-Frequency (ADMX-HF) is a sister experiment to ADMX  based at Yale U. It is a collaboration of Yale U., U. of Colorado/JILA, UC Berkeley, U. of Florida and LLNL. It is primarily sponsored by the National Science Foundation (NSF) with additional DOE Office of Science contributions through LLNL participation. The experiment is conceptually similar to ADMX except that it has a smaller volume (5 inch bore) but higher field (9.4 T) magnet. It will primarily begin to search for axions in the 4-6 GHz range (or 16-24 $\mu$eV) but likely will only be able to reach KSVZ sensitivity given that the cavity volume is $\sim 1$\% that of ADMX. 

\section{Discussion and Conclusion}
\begin{figure*}[tbh]
  \includegraphics[angle=0,width=0.95\textwidth]{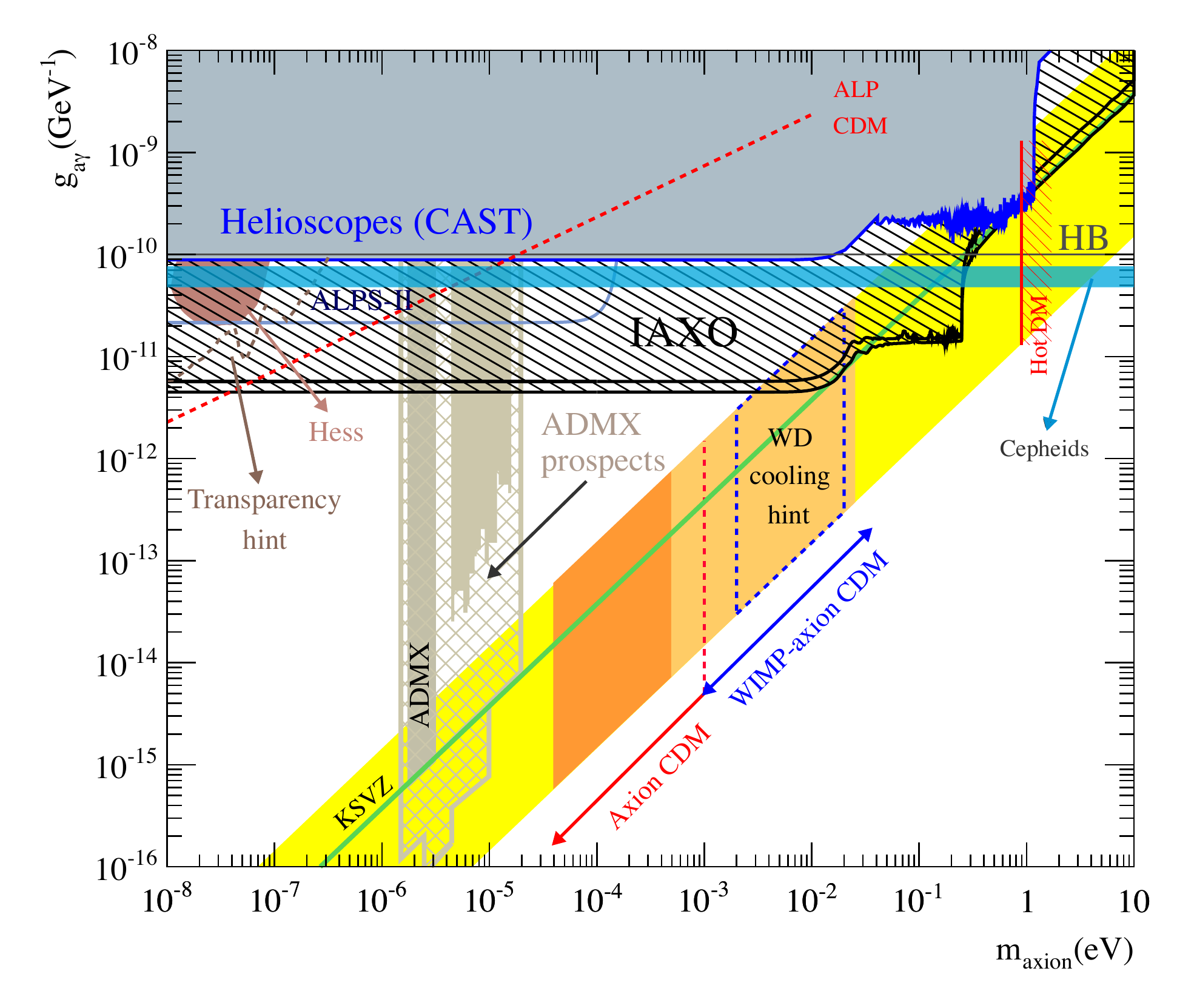}
\caption{Summary of results, prospects, and hints in the axion/ALP parameter space.
The QCD motivated models for axions lay in the yellow diagonal band. 
In orange are the regions of cosmological interest. 
Axions in those areas may be part of the CDM in the universe. The region between $ m_a=10^{-5} $ and $ 10^{-3} $ eV is often referred to as the ``classic" CDM region.
The different colors indicate different Dark Matter scenarios (see caption of Fig. 1 in \cite{Irastorza13} for a full explanation). 
In the case of ALPs, the CDM region is represented by the space below the dashed red line.
The prospects of ADMX are shown in the hatched brown region and the expectations of IAXO in the black hatched region. 
The astrophysical bounds from HB and massive stars are labeled ``HB'' and ``Cepheids'' respectively.
The low-mass region motivated by the transparency hints is above the dashed gray line.
}
\label{Fig:overview}
\end{figure*}
In this note we have briefly summarized the status of the helioscope and haloscopes experiments in probing the axion/ALP-photon coupling $ \g $, and discussed the phenomenological bounds from He-burning stars. 

A summary of the results is shown in Fig.~\ref{Fig:overview}.
The QCD motivated region is shown as a yellow diagonal band (the width is somewhat arbitrary~\cite{Irastorza13}). 
The green line within the band represents the KSVZ axion model~\cite{Kim:1979if,Shifman:1979if}.

The astrophysical bound from Horizontal Branch stars is shown by the  line labeled
``HB''. The excluded region lies above the line. The blue band below it,  labeled ``Cepheids'', indicates the corresponding bound from the blue loop stage of massive stars. The width indicates the astrophysical uncertainties, as discussed in~\cite{Friedland:2012hj}, with the values above the band being conservatively excluded.
These constraints are essentially mass-independent in the region shown in the graph (see section \ref{sec:astro}) and hence appear as horizontal lines.


For $ m_a<0.02 $ eV, the astrophysical bounds represent a slight improvement over the CAST results, which is currently  the most powerful axion helioscope.
Even more importantly, for larger axion masses, they have no laboratory analogues. It is in this region that they overlap with the QCD-motivated band.

The low mass region explored by CAST is also of physical interest because of a few phenomenological hints which point in the direction of axion-like particles. 
In fact, the region is also being analyzed by other experiments, not discussed in this note.
Notably, Hess has excluded a narrow mass region down to a coupling of about $ 0.5\times 10^{-10} $ GeV$ ^{-1} $ and ALPS II plans to explore that region further in a broader range of masses~\cite{Baker:2012nq}.

As clear from the figure, a next generation axion helioscope as the proposed IAXO would have 
an impressive capability of pushing the ALP-photon coupling to lower values, while also enlarging the mass range.

IAXO is a fourth-generation axion helioscope concept, envisioning the construction of a dedicated magnet and reflective x-ray optics to dramatically increase its sensitivity compared to the previous generation helioscopes. 
Preliminary studies (see, e.g., \cite{Irastorza13}) predict that IAXO would be able to explore the region of masses $ m_a<20 $ meV down to $ \g \simeq$ a few $ 10^{-11} $ GeV$ ^{-1} $
and would shed light on the QCD motivated region at the high-mass end of the exclusion diagram shown. 
Finally, IAXO has also the potential to serve as multi-purpose facility for generic axion and ALP research in the next decade~\cite{Isern:2013lwa}. 

ADMX has probed the small coupling region to values inaccessible to any other experiments. 
The new ADMX-HF will expand considerably the mass range explorable in the near future. 
These two experiments are unique for exploring the well-motivated region of parameters in which the QCD axion can be Dark Matter.

\begin{acknowledgments}
At Los Alamos National Laboratory, this work was support by the Department of Energy, Office of High Energy Physics, and by the LANL LDRD program.
Part of this work was performed under the auspices of the U.S. Department of Energy by Lawrence Livermore National Laboratory under Contract DE-AC52-07NA27344 and under the auspices of the National Science Foundation, under Grants No. PHY-1067242 and No. PHY-1306729. 
\end{acknowledgments}

\bibliography{axionbib}

\end{document}